\documentclass[aps,prb,twocolumn, groupedaddress,superscriptaddress,showpacs]{revtex4}
\usepackage{graphicx,amsmath}
\usepackage{amssymb}
\usepackage{nicefrac}

\begin{document}
\title{Density-Functional Theory and Tight-Binding Studies of the Geometry of Hydrogen Adsorbed on Graphynes}

\author{Hunpyo Lee}
\email[]{hplee@sissa.it}
\affiliation{School of Physics, Konkuk University, Seoul 143-701, Korea}
\affiliation{CNR-IOM-Democritos National Simulation Centre and International
School for Advanced Studies (SISSA), Via Bonomea 265, I-34136, Trieste, Italy}
\author{Jahyun Koo}
\affiliation{School of Physics, Konkuk University, Seoul 143-701, Korea}
\author{Massimo Capone}
\affiliation{CNR-IOM-Democritos National Simulation Centre and International
School for Advanced Studies (SISSA), Via Bonomea 265, I-34136, Trieste, Italy}
\author{Yongkyung Kwon}
\affiliation{School of Physics, Konkuk University, Seoul 143-701, Korea}
\author{Hoonkyung Lee}
\email[]{hkiee3@gmail.com}
\affiliation{School of Physics, Konkuk University, Seoul 143-701, Korea}

\date{\today}

\begin{abstract}
Using density-functional theory and a tight-binding approach we 
investigate the physical origin of distinct favourable geometries of adsorbed
hydrogen atoms in various graphyne structures, and the relation with
electronic properties. In particular, H atoms are adsorbed in-plane for 
$\alpha$-graphyne, and they assume an oblique configuration in all other 
graphynes, including 6,6,12-graphyne. The origin of different configurations is
identified by means of a simple tight-binding model and it is
controlled by the tuning of the hopping 
between sp$^2$-bonded C atoms and sp-bonded C 
atoms hybridized with the H atoms. 
We discuss in details how the geometry change of 
the attached H atom tunes the electronic properties like energy gap. 
\end{abstract}



\maketitle

\section{Introduction\label{Introduction}}

The synthesis of graphene, a two-dimensional atomic layer of carbon
atoms on a honeycomb lattice, has generated a new area of
condensed-matter
physics in which basic physics associated with the existence of Dirac cones is intertwined with a huge potential
for applications such as 
electronic devices, hydrogen storage materials, and lithium-ion battery 
materials~\cite{Novoselov2004,Elias2009,Zhang2005,Neto2009}.

Device-oriented applications of graphene are based on the possibility
to open and tune a band gap from the 
semimetallic Dirac states~\cite{Choi2010,Pereira2009,Mohr2009}. 
This can be realized by breaking the
sublattice symmetry or the chiral
symmetry~\cite{Zhou2007,Kim2008,Son2006,Lee2011}, and by
chemical functionalization of adsorbing hydrogen or fluorine on
graphene~\cite{Choe2010,Boukhvalov2008}. 
Even if the functionalization could in principle open the
gap, the adsorbed atoms tend to segregate in clusters, strongly
limiting the applicative potential of these phases.

\begin{figure}[htbp]
\includegraphics[width=8.1cm,height=8.5cm]{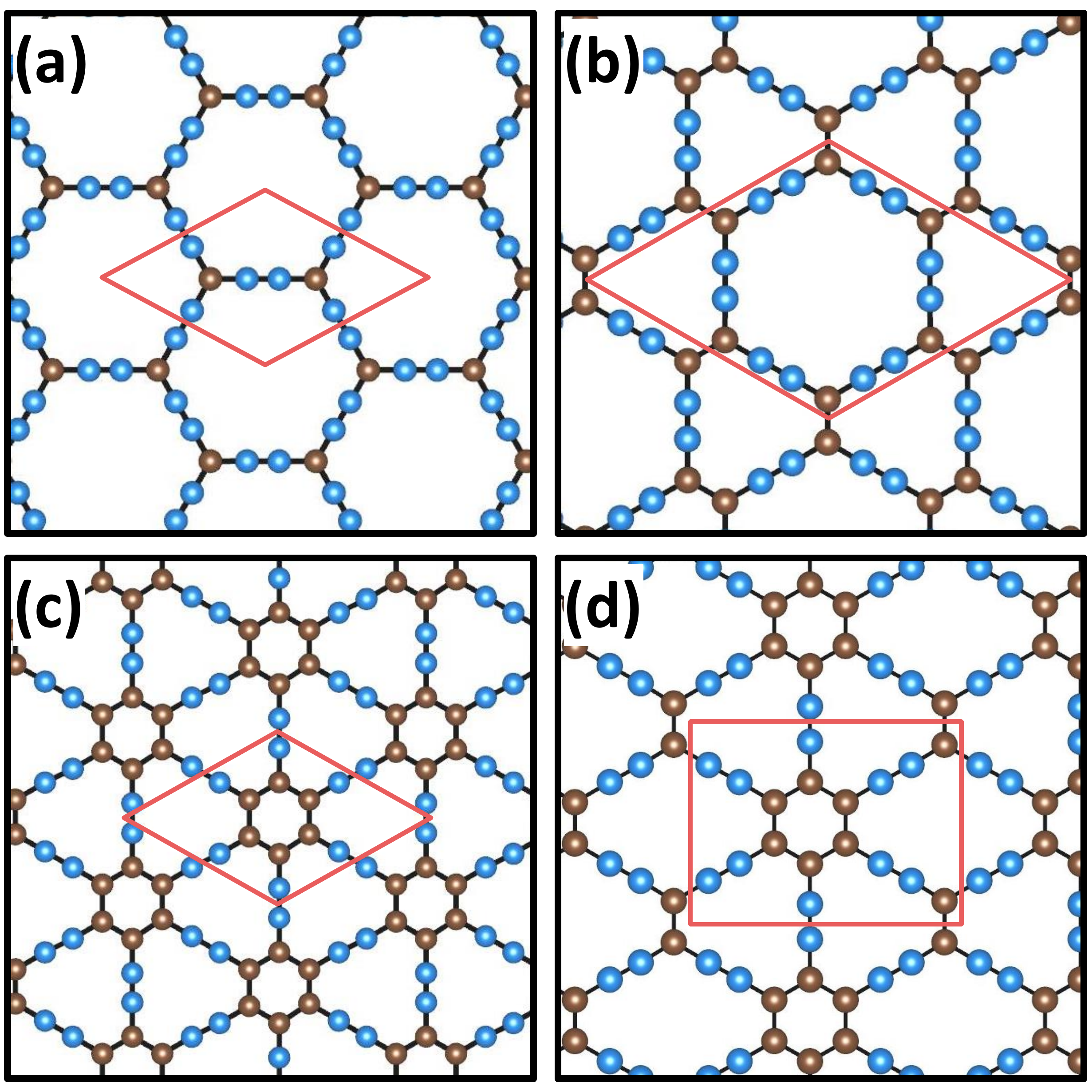}
\caption{(Colour online) (a) Schematic pictures of (a) $\alpha$-, (b) $\beta$-, 
(c) $\gamma$-, and (d) 6,6,12-graphynes. 
Red quadrlaterals indicate unit cell for density 
functional theory and tight-bind calculations. The brown 
and blue circles present the sp$^2$-bonded C atoms and sp-bonded
C atoms hybridized with H atoms, respectively.} \label{Fig1}
\end{figure}

A different route to gap-tuning in graphene-like structures is recently 
offered by graphyne~\cite{Malko2012,Kim2012,Koo2013,Liu2012,Popov2013}. 
Graphynes can be ideally
obtained inserting carbon atoms with triple bonds (C$\equiv$C)
between pairs of sp$^2$-bonded carbons in graphene, forming a 
sp-sp$^2$ hybrid carbon network on a 
two-dimensional hexagonal lattice~\cite{Baughman1987}. 
Different graphyne structures have been proposed 
according to position and number of added 
C$\equiv$C units. In Figs.~\ref{Fig1} (a), (b), (c), and (d) 
we show the so-called
$\alpha$-, $\beta$-, $\gamma$-, and 6,6,12-graphyes.
Density functional  theory (DFT) calculations predict a semimetallic
state with Dirac cones in $\alpha$- and $\beta$-graphyne 
and semiconductor with a band gap of 0.47 eV in 
$\gamma$-graphyne~\cite{Kim2012}.  More recently, a semimetal states
with double Dirac cones has been predicted  for 6,6,12-graphyne~\cite{Malko2012}.


Just like in graphene, one can in principle open and tune a band gap
in the different graphyne structures. 
Some of us have shown by means of DFT calculations that adsorbed
hydrogen atoms prefer different geometries according to the type of
graphyne~\cite{Koo2013}, leading in turn to different electronic properties. 
For instance, H atoms in $\alpha$-graphyne (C$_1$H$_{0.75}$), where
each sp-bonded C atom accommodate one H atom, 
prefer an in-plane configuration and the electronic structure remains
semimetallic regardless of the attached H atoms, while H atoms 
adsorbed on $\gamma$-graphyne prefer an oblique configuration with
respect to the plane, and the  energy band gap dramatically widens to
2.19 eV from the 0.47 eV of the pure compound~\cite{Koo2013}. 

These results confirm the possibility to tune the electronic properties
via hydrogenation also in view of applications. As opposed to
graphene, H atoms are able to diffuse in graphynes, opening a path
towards application opportunities in devices and and hydrogen storage.

\begin{figure}[htbp]
\includegraphics[width=0.475\textwidth]{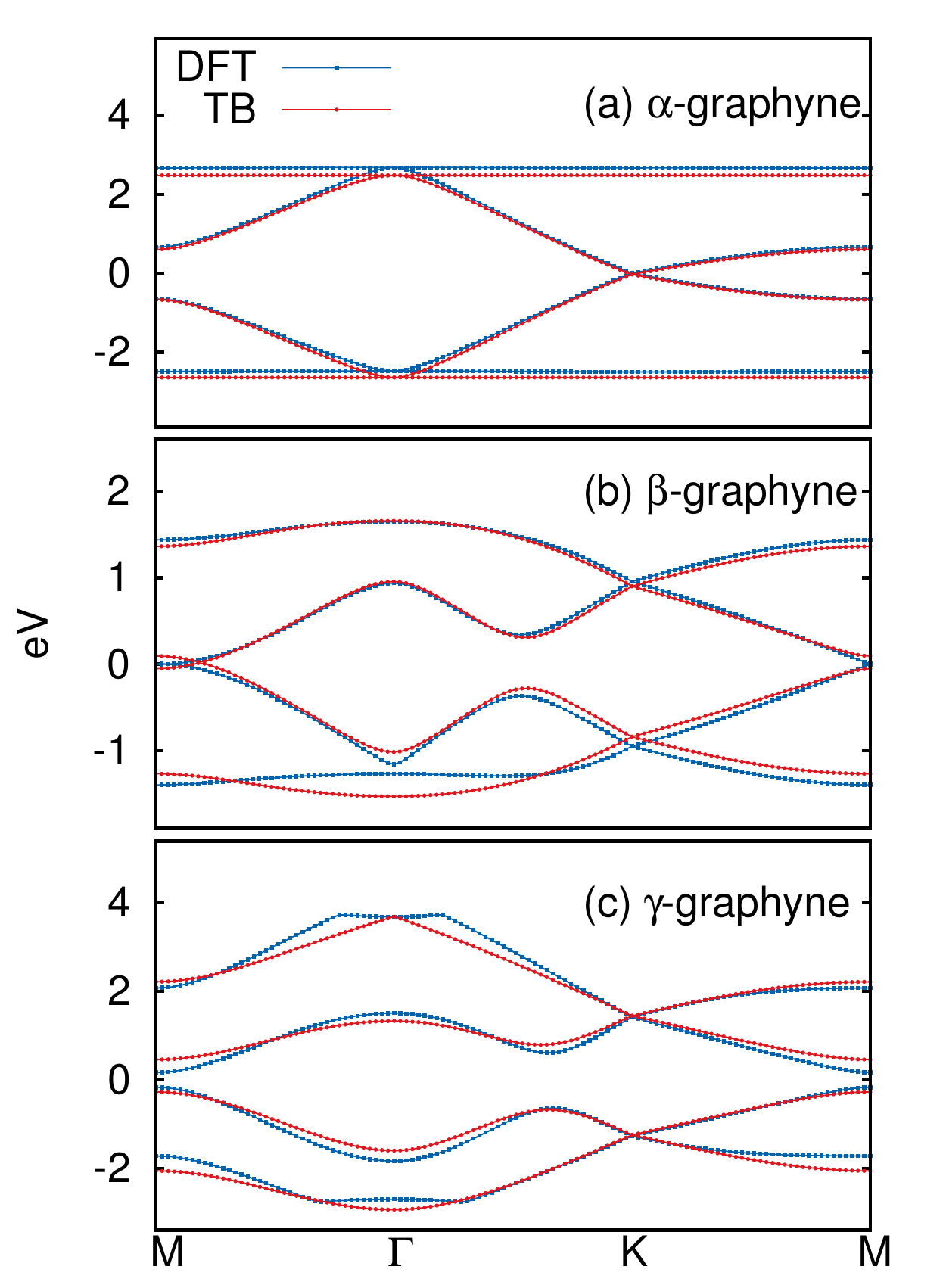}
\caption{(Colour online) Bandstructures along the M-$\Gamma$-K-M
directions of (a) $\alpha$-, (b) $\beta$-, and (c) $\gamma$-graphynes 
calculated by 
DFT and tight-binding calculations with
in-plane H atoms.
The energies are in unit of eV and referred to the Fermi energy. 
$\alpha$ and $\beta$-graphyne behave as semimetal, while the $\gamma$-graphyne 
is an insulator with a gap of 0.34 eV.} \label{Fig2}
\end{figure}

Here we performed DFT calculations with more k-points  
than former DFT ones 
for hydrogenated $\alpha$- and $\gamma$-graphynes, 
and we newly carried out DFT calculations for hydrogenated $\beta$- and 
6,6,12-graphynes. In order to identify the physical origin of these
different arrangements, we use a tight-binding approach based on the
DFT results, and we connect the angle between the C-H bond and the
graphyne plane (i.e., the geometry of the adsorbed H) to the hopping between 
neighbouring sp$^2$-bonded C and and sp-bonded C with attached H. 

Our tight-binding calculations confirms that in $\alpha$-graphyne the
energy is minimized by the hopping which corresponds to in-plane H
atoms, while for $\beta$- and $\gamma$-graphynes tuning the
above-defined hopping parameter leads to a lower energy for a finite
angle, identifying the physical mechanism behind the different
geometrical arrangements.

Finally, we find that adsorption of one H atom per C in 6,6,12-graphyne turns
the semimetal into a semiconductor and we discuss the geometric and electronic properties of 
hydrogenated compound. 

The paper is organized as follows: in Sec.~\ref{DFTmethod} and ~\ref{TBmodel}, 
we describe the computational detail of the electronic 
structure calculations and the tight-binding model, respectively. 
In the first and second parts of the Sec.~\ref{Results}, 
we discuss the DFT as well as TB results of hydrogenated 
$\alpha$-, $\beta$-, and $\gamma$-graphynes in in-plane and optimal 
configurations of the attached H 
atom in C atoms, respectively. 
In the last part of the Sec.~\ref{Results}, we analyse 
geometrical and electronic structures of the optimal 6,6,12-graphyne using the 
DFT calculations. In Sec.~\ref{Conclusions}, we summary our findings.

\section{COMPUTATIONAL DETAILS}\label{DFTmethod}

Our calculations were performed using a first-principles method based on density 
functional theory as implemented in the Vienna Ab-initio Simulation Package 
with a projector-augmented-wave (PAW) method~\cite{Kohn1965,Kresse1999}. 
The exchange correlation energy functional was used with the generalized 
gradient approximation in the Perdew–Burke–Ernzerhof scheme, and the 
kinetic energy cutoff was set at 400 eV. Our model graphyne systems were a 1 
$\times$ 1 hexagonal cell. 
A geometrical optimization of H-adsorbed $\alpha$-graphyne was carried out 
within a fixed 1 $\times$ 1 cell 
obtained from the equilibrium lattice constant of the isolated graphyne until 
the Hellmann-Feynman force acting on each atom was less than 0.01 eV/A$^\circ$. The 
first Brillouin zone integration was done using the Monkhorst–Pack scheme. 
A 8 $\times$ 8 $\times$ 1 k-point sampling was done for the 
1 $\times$ 1 graphyne cell. To remove spurious 
interactions between image structures due to periodic calculations, a vacuum 
layer of 12 A$^\circ$ was taken in each of all nonperiodic directions.

\section{TIGHT-BINDING MODEL}\label{TBmodel}
We performed DFT calculations on hydrogenated 
$\alpha$-, $\beta$-, and $\gamma$-graphyne in the configuration where each sp-bonded C atom 
accommodates one {\it in-plane} H atom. 
We chose $N_c=8$, 18, and 12-sites in the unit 
cell for $\alpha$-, $\beta$-, and $\gamma$-
graphynes, respectively and we optimized the lattice parameters.
The blue circles in Figs.~\ref{Fig1} (a), (b), and (c) 
show the position of sp-bonded C atoms hybridized with the H atoms 
on $\alpha$-, $\beta$-, and $\gamma$-
graphynes with chemical
formula C$_1$H$_{0.75}$, C$_1$H$_{0.67}$, and C$_1$H$_{0.5}$, 
respectively. We obtained that $\alpha$- and $\beta$-graphynes are semimetal and $\gamma$-graphyne is semiconducting with 
a band gap of 0.34 eV. These results are qualitatively and also
quantitatively similar to those of pure graphyne.

Based on these DFT results we have derived a tight-binding Hamiltonian.
Our parameterization neglects the hybridization effects between the sp-bonded C atoms and 
the attached H atoms because the DFT results do not show significant
differences in the electronic structures of pure and hydrogenated
graphynes. Consequently we assume that the bonding 
configuration among the p$_z$ orbitals of different sites still form 
$\pi - \pi^*$  bands, like pure cases without H atoms.

Our tight-binding Hamiltonian has therefore the form
\begin{equation}
H= -\sum_{<ij>,\sigma} t_{mm'} (c_{im,\sigma}^{\dagger}c_{jm',
\sigma} + \text{H.c}) -\mu_{m'} \sum_{im',\sigma} n_{im',\sigma},
\label{Eq:hamiltonian}
\end{equation}
where the site index ($i$ or $j$) runs over all the carbon atoms,
which can be either sp$^2$-bonded (labelled as ``2'') and sp-bonded
(labelled as ``1'') hosting a H atom. The two carbon atoms have
different local energies $\mu_{m}$ with $m =1,2$. The hopping is
restricted to nearest neighbours, but different hopping amplitudes are
associated to bonds connecting (i) two sp-bonded C atoms
($t_{11}$), (ii) one sp$^2$-bonded carbon with one sp-bonded
hydrogenated carbon ($t_{12}$) and (iii) two sp$^2$-bonded C atoms
($t_{22}$). The hoppings and local energies are obtained simply by
fitting the DFT bands.

\section{RESULTS}\label{Results}

\subsection{In-plane absorption on hydrogen atoms}

As mentioned above, we start from DFT calculations with in-plane H
atoms. The hopping parameters $t_{mm'}$ and 
on-site energies $\mu_{m'}$ that we obtain by fitting results for
$\alpha$-, $\beta$-, and $\gamma$-graphynes are presented in Table 1
(the energy unit is eV). 
\begin{table}[htbp]
\begin{tabular}{l*{6}{c}r}
\hline\hline
&& Graphynes & $t_{22}$ & $t_{12}$ & $t_{11}$ & $\mu_1$ & $\mu_2$ \\
\hline
&& $\alpha$ graphyne & 0.0 & 2.458 & 2.561 & -0.0781 & 0.0987 \\
&& $\beta$ graphyne  & 3.1772 & 2.6670 & 2.4369 & -0.2046 & 0.11081\\
&& $\gamma$ graphyne & 2.3964 & 2.3833 & 3.1987 & 0.4844 & -0.4491\\
\hline\hline
\end{tabular}
\caption{$t_{22}$ is the hopping between sp$^2$-bonded C atoms, 
$t_{12}$ is the hopping of between the sp$^2$-bonded C atoms and 
sp-bonded C atoms with adsorbed H atoms, 
and $t_{11}$ is the hopping between sp-bonded C atoms 
with the absorbed H atom. $\mu_1$ and $\mu_2$ are on-site energies in C and
C atoms with the attached H atom, respectively.}
\label{parameter}
\end{table}
Since $\alpha$-graphyne is topologically equivalent to graphene, we
can estimate an effective hopping between carbon atoms in the graphene
honeycomb lattice. We obtain a value of 0.7 eV, which is slightly smaller than that of 
graphene and pure $\alpha$-graphyne without the absorbed H
atoms~\cite{Liu2012}.

In principle one may expect that both $\beta$- and $\gamma$-graphynes
should become insulating with a charge-density-wave ordering because
of the broken symmetry between the two sublattices, as it has been
observed recently in quasi-neutral molecular graphene 
with an additional CO molecule~\cite{Gomes2012}. Our calculations
confirm this expectation only for $\gamma$-graphyne, while 
the $\beta$-graphyne configuration remains semimetallic.

\begin{figure}[htbp]
\includegraphics[width=0.5\textwidth]{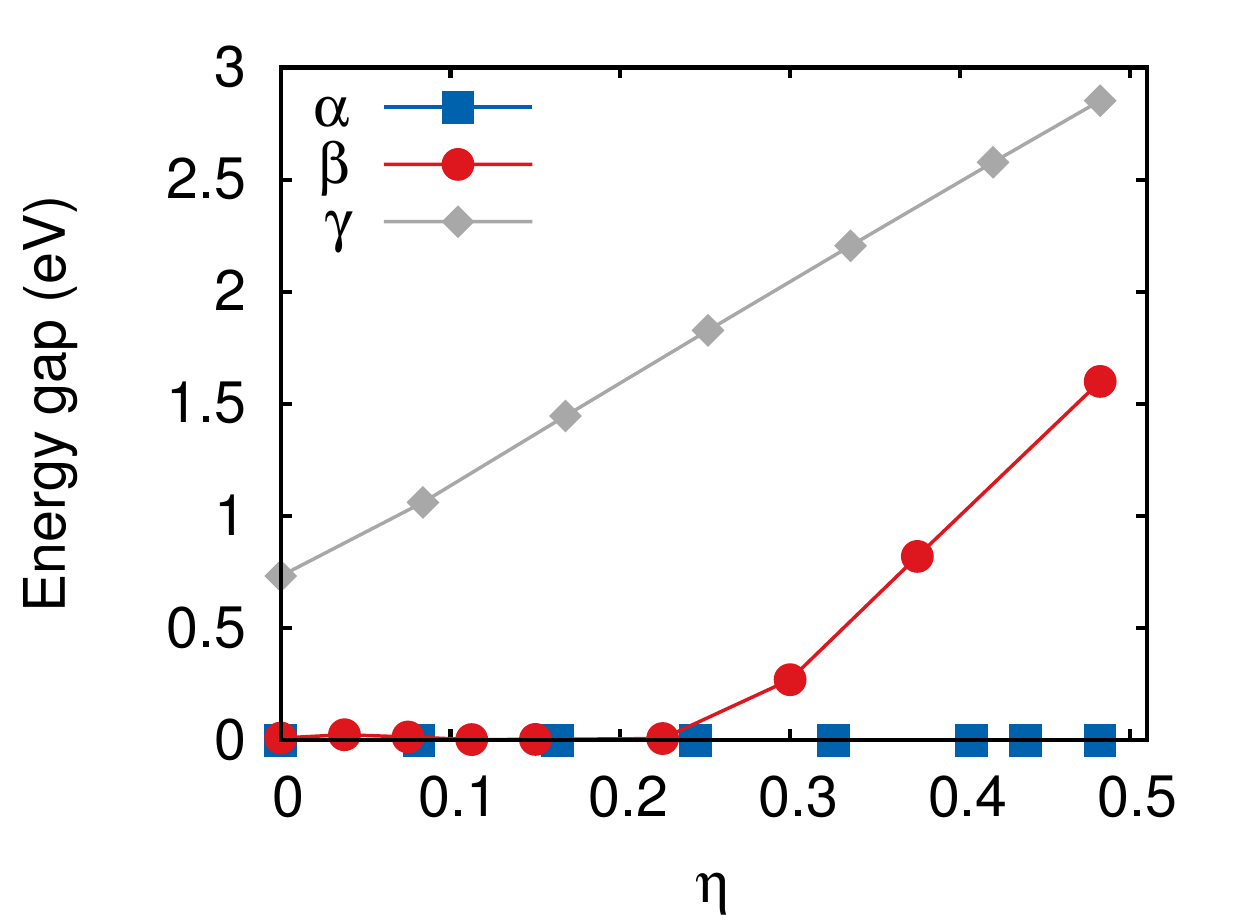}
\caption{(Colour online) The calculated tight-binding  energy gap (Energy gap)
as a function of $\eta$ for $\alpha$-, $\beta$-, and $\gamma$-graphynes.
The renormalization parameter $\eta$ is related to 
geometrical configuration of H atoms attached in sp-bonded C atoms.} 
\label{fig:gap}
\end{figure}

In order to account for why the Dirac cone is still present in $\beta$-graphyne 
but absent in $\gamma$-graphyne, we write the simple $2 \times 2$
Hamiltonian on honeycomb structure with different 
hopping $\tau_1$ and $\tau_2$ along the inequivalent directions.
The Hamiltonian is given as
\begin{equation}
H =
 \begin{pmatrix}
  0 &  f(\bf{k}) \\
  f^{*}(\bf{k}) &  0
 \end{pmatrix}
\end{equation} 
with $f(\bf{k}) = \tau_1 e^{ik_x}+\tau_2 (e^{i(-\frac{k_x}{2}+\frac{\sqrt{3}k_y}{2})}+e^{i(-\frac{k_x}{2}-\frac{\sqrt{3}k_y}{2})})$ and
the lattice constant between sites is set into one for simplicity. The
2$\times$2 matrix structure is associated to inequivalent lattice
sites on the two sublattices. 
It is easy to verify that a Dirac dispersion is present only if $\vert \tau_1 \vert$ is smaller than $\vert 2 \tau_2 \vert$~\cite{Kim2012}. 
The DFT results can be therefore rationalized in terms of the
effective hopping between C atoms on the sites of the honeycomb
lattice. The semimetallic state in $\beta$-graphyne is therefore
explained by an effective $\tau_1^{'}$ smaller than 
$\vert 2  t_{22} \vert$, while the charge-density-wave insulating state
with 
a band gap of 0.23 eV of $\gamma$-graphyne is a consequence of a
larger effective hopping on the ideal honeycomb lattice, which
satisfies$\vert \tau_1^{'}\vert > \vert 2 t_{22}\vert$.

\begin{figure}[htbp]
\includegraphics[width=0.475\textwidth]{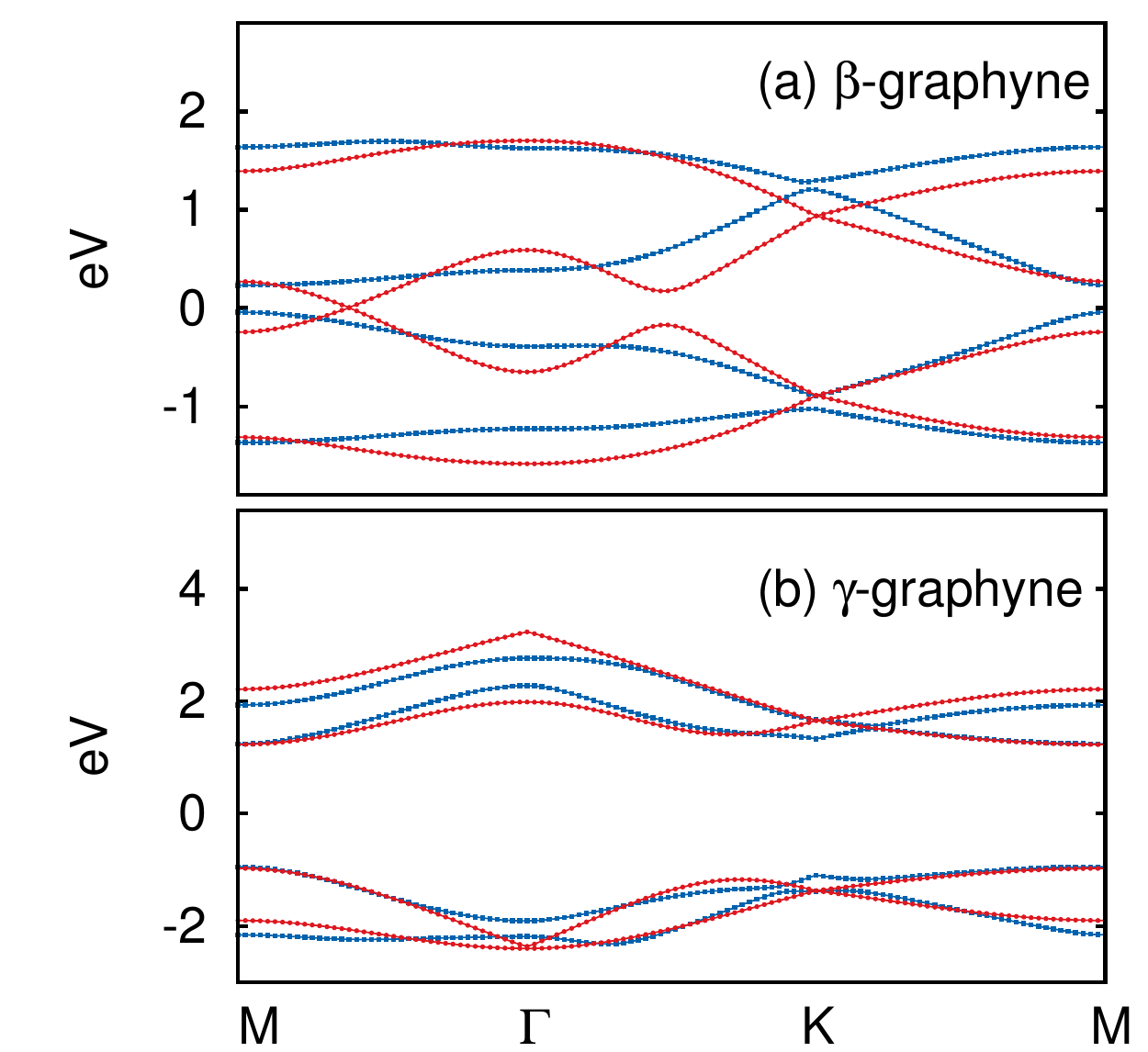}
\caption{(Colour online) Band structures along the M-$\Gamma$-K-M directions of 
(a) $\beta$-, and (b) $\gamma$-graphynes calculated by DFT in the optimal 
configuration of attached H atoms and
tight-binding with the value of $t_{12}$ which reproduces 
the DFT gap at the M-point of the Brillouin zone, namely 2.41 eV ($\beta$) and
1.58 eV ($\gamma$). Energies in eV are referred to the Fermi level.} \label{fig4:optimal_band}
\end{figure}

\subsection{Optimal configurations of the attached H atoms}

DFT calculations show that the 
in-plane configuration of the attached H atoms is stable only in
semimetallic $\alpha$-graphyne, while $\beta$- and $\gamma$-graphynes
choose a finite-angle configuration with a band gap of 0.27 eV 
and 2.19 eV, respectively. Here we elaborate on the physical origin of
the different configurations. 
In a very simple picture, if the H atom is in the ``oblique''
configuration, the p$_z$-p$_z$ hybridization between sp$^2$-bonded 
C without
H atom and sp-bonded C 
with H atom would be reduced, because the H atom attracts the $\pi$ orbital 
in sp-bonded C atom. 
Therefore, a finite angle between the C-H bond and the graphyne plane
reflects in a reduced hopping element $t_{12}$. We denote with 
$t_{12}^{\text{inplane}}$ the DFT-derived value for an in-plane H
adsorption and we perform calculations tuning the value of $t_{12}$ to
effectively take into account a finite angle. 
We define a renormalization parameter 
\begin{equation}
\eta=\frac{t_{12}^{\text{inplane}} - t_{12}}{t_{12}^{\text{inplane}}},
\end{equation}
so that $\eta=0$ and $1$ describe the in-plane configuration  and
the complete separation between the hexagons, respectively.

\begin{figure}[htbp]
\includegraphics[width=0.5\textwidth]{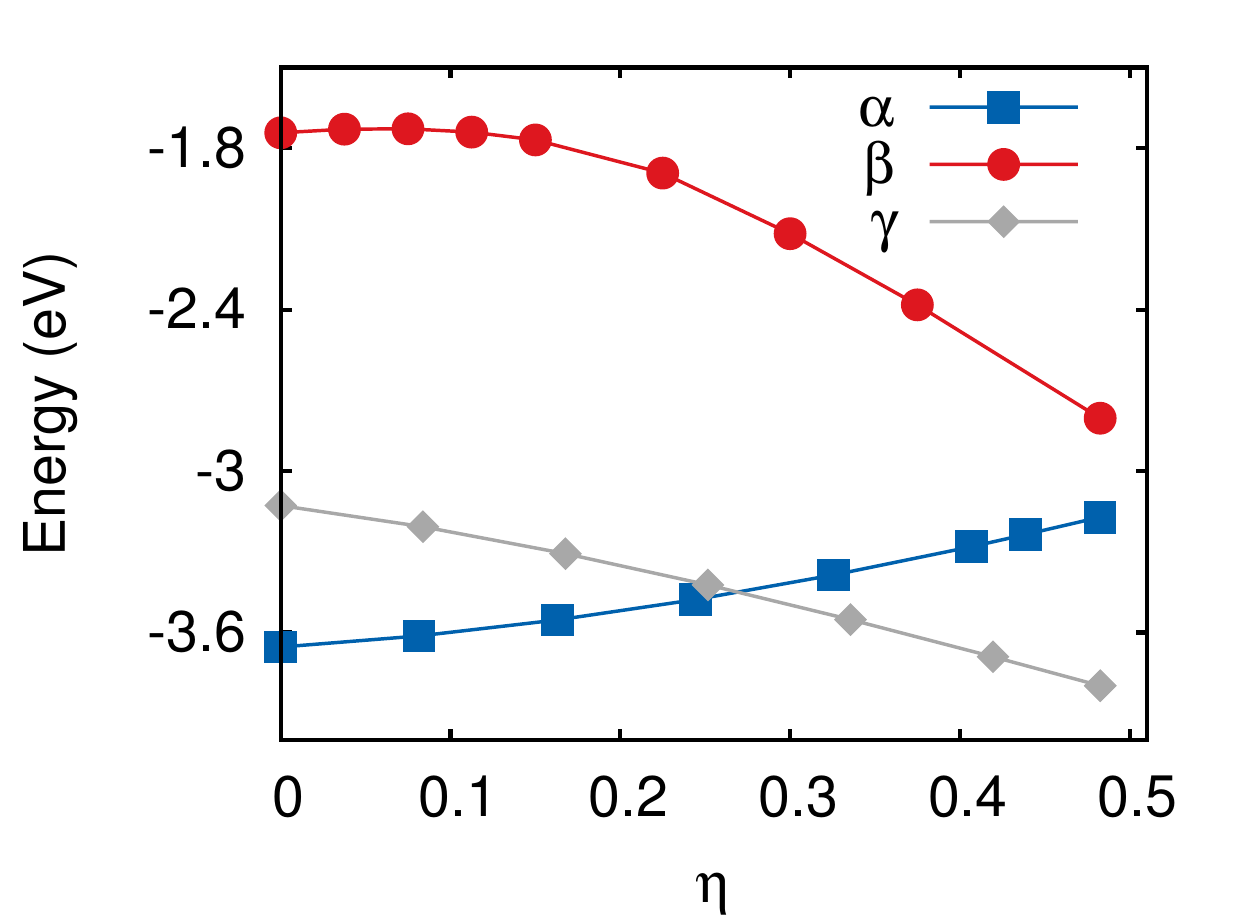}
\caption{(Colour online) Band energy (Energy) as a 
function of $\eta$ for $\alpha$-, $\beta$-, and $\gamma$-graphynes in 
the tight-binding model. The renormalization parameter $\eta$ is related to the
geometrical configuration of H atoms attached in sp-bonded C atoms.} \label{fig:energy}
\end{figure}

In Fig.~\ref{fig:gap} we plot the energy gap as a function of $\eta$ for
our tight-binding representations of $\alpha$-, $\beta$- and $\gamma$-graphyne.
For $\alpha$-graphyne the system remains semimetallic increasing
$\eta$. On the other hand, in
$\gamma$-graphyne the gap increases linearly with $\eta$, reaching the
DFT value of 2.19 eV for $\eta = 0.33$, which corresponds to $t_{12} =
1.58$ eV. For $\beta$-graphyne, the gap opens for $\eta = 0.23$. 

In Figs.~\ref{fig4:optimal_band} (a) and (b)
we compare the DFT bandstrucures of $\beta$- and $\gamma$-graphyne with those 
obtained in the tight-binding model for the value of $t_{12}$ which reproduces 
the DFT gap at the M-point of the Brillouin zone, namely 2.41 eV ($\beta$) and
1.58 eV ($\gamma$). 
This comparison measures to which extent our simple idea to model the
geometrical character of the adsorbed H atoms reproduces the actual
DFT calculations. 
The agreement is extremely good over the whole Brillouin zone for
$\gamma$-graphyne, while some discrepancies are visible in
$\beta$-graphyne.
We conjecture that the discrepancy is most likely due to next nearest neighbour 
hopping term, which we neglect here in order to keep the model as
simple and transparent as possible. 

Finally, we plot the band energies as a function of $\eta$ including
the four lowest-energy bands around Fermi level at each graphyne 
in Fig.~\ref{fig:energy}. 
For $\alpha$-graphyne the energy is minimized by $\eta=0$, confirming
that the adsorbed H atoms prefer to sit in the plane, as in DFT.
On the other hand, for $\gamma$-graphyne the energetically favoured
configuration is $\eta = 1$, which corresponds to separated hexagons,
while in the actual DFT calculation a given configuration with oblique
H configuration is stabilized by lattice deformation and
hybridization between C and H atoms, which are not included in our
very simple tight-binding model.  
As mentioned above, if we compare energy band, 
a coefficient $\eta = 0.33$ reproduces 
the correct gap amplitude and the overall bandstructure.

\subsection{Hydrogenated 6,6,12-graphyne}

In this section we investigate within DFT the electronic properties of
hydrogenated 6,6,12-graphyne  with a concentration of C$_1$H$_{0.56}$, 
where each sp-bonded C atoms hosts one H atom. Recent calculations have revealed a double Dirac cone for
pure 6,6,12-graphyne~\cite{Malko2012}. 
We start our analysis by optimizing the H positions within the
in-plane configuration. Here the Dirac cones are found to be replaced with a small band gap opening of 0.03 eV as a
consequence of the broken sublattice symmetry in x+y and x-y
directions.

Next, we find the optimal lattice structure releasing the in-plane
constraint for the H atoms. The optimal lattice structure with 
an oblique configuration of the H atoms attached in sp-bonded C atom
and the  
electronic structure with an energy gap of 1.17 eV are shown in upper and lower 
parts of Fig. 6, respectively. 
\begin{figure}[htbp]
\includegraphics[width=0.49\textwidth]{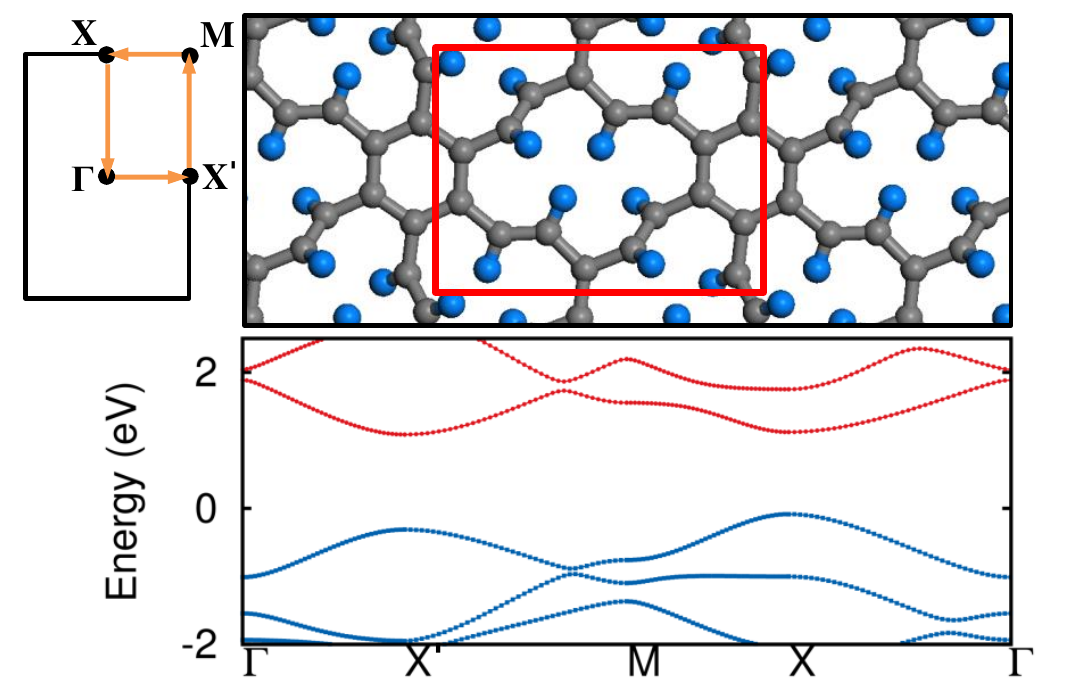}
\caption{(Colour online) (Upper panel) Atomic structures of 
hydrogenated 6,6,12-graphyne with a concentration of C$_1$H$_1$, 
where each sp-bonded C atoms 
include one H atom. The oblique configuration of H atom absorbed in sp-bonded C 
atom is energetically favoured. (Lower panel) The DFT bandstructure of 
hydrogenated 6,6,12-graphyne with a concentration of C$_1$H$_{0.56}$. 
The system is a semiconductor with an energy gap of 1.17 eV.} 
\label{fig:6612result}
\end{figure}
The attached H atoms in the oblique configuration determine a larger
symmetry breaking potential with respect to the in-plane configuration in the 6,6,12-graphyne. 
Therefore, the energy gap becomes much larger and the oblique state is
energetically stabilized, just like in $\gamma$-graphyne.

\section{SUMMARY}\label{Conclusions}

In conclusion, we have explored the geometrical configurations of H atoms 
absorbed on sp-bonded C atoms of $\alpha$-, $\beta$-, $\gamma$-, 
and 6,6,12-graphynes and the consequent electronic properties. In the
latter case we show for the first time that adsorption of hydrogen
removes the Dirac cones of pure 6,6,12-graphyne leading to a
semiconductor with a gap of 1.17 eV which strongly depends on the H atom configuration.

Using DFT calculations, we first studied the electronic properties 
of the cases with in-plane configurations of the adsorbed H atoms in 
$\alpha$-, $\beta$-, and $\gamma$-graphynes. 
And then, we established a simplified tight-binding model with 
lattice parameters based on the in-plane configurations 
of attached H atoms. Starting from the tight-binding model, we
mimicked the p$_z$-p$_z$ hybridization associated to an off-plane
oblique configuration by means of a tuned hopping parameter.    
This simple picture allows to understand why the in-plane configuration of H 
atom is stable in 
$\alpha$-graphyne, while oblique configurations are favoured
in $\beta$- and $\gamma$-graphynes. 
Moreover, we find that these different geometrical 
configurations strongly affect the opening and the size of the energy
gap, suggesting possible directions to control it in view of applications.

\section{ACKNOWLEDGMENTS}
H.L, J.K, Y.K, and H.L are supported 
by the Basic Science Research Program (Grant No. KRF-2012R1A1A1013124) through 
the National Research Foundation of Korea, funded by the Ministry of Education, 
Science and Technology. H.L and M.C are supported by ERC/FP7 
through the Starting Independent Grant ”SUPERBAD”, Grant Agreement No. 240524.


\end{document}